\documentclass[twocolumn]{aastex61}
\usepackage{amsmath}
\usepackage{todonotes}

\defcitealias{marble10}{M10}
\usepackage{color}
\shorttitle{Molecular cloud N66}
\shortauthors{Naslim et al.}

\begin{document}

\title{ALMA reveals a cloud--cloud collision that triggers star formation in N66N of the Small Magellanic Cloud}
%\correspondingauthor{Naslim Neelamkodan}
%\email{naslim.n@uaeu.ac.ae}

%\affiliation{National Centre for Radio Astrophysics, Tata Institute of Fundamental Research, PO Box 3, Pune, 411007, India}
\author[0000-0001-8901-7287]{Naslim~N}
\affiliation{Department of Physics, United Arab Emirates University, Al-Ain, UAE, 15551.}
\author[0000-0002-2062-1600]{Kazuki Tokuda}
\affiliation{Department of Physical Science, Graduate School of Science, Osaka Prefecture University, 1-1 Gakuen-cho, Sakai, Osaka 599-8531, Japan}
\affiliation{National Astronomical Observatory of Japan, National Institutes of Natural Science, 2-21-1 Osawa, Mitaka, Tokyo 181-8588, Japan}
\author{Susmita Barman}
\affiliation{School of Physics, University of Hyderabad, Prof. C. R. Rao Road, Gachibowli, Telangana, Hyderabad, 500046, India}
\author{Hiroshi Kondo}
\affiliation{Department of Physical Science, Graduate School of Science, Osaka Prefecture University, 1-1 Gakuen-cho, Sakai, Osaka 599-8531, Japan}
\author{Hidetoshi Sano}
\affiliation{National Astronomical Observatory of Japan, National Institutes of Natural Science, 2-21-1 Osawa, Mitaka, Tokyo 181-8588, Japan}
\author{Toshikazu Onishi}
\affiliation{Department of Physical Science, Graduate School of Science, Osaka Prefecture University, 1-1 Gakuen-cho, Sakai, Osaka 599-8531, Japan}

%\correspondingauthor{Naslim Neelamkodan} 
%\email{naslim.n@uaeu.ac.ae}
%\author[0000-0002-0844-6563]{Naslim. N}
%\affiliation{Department of physics, United Arab Emirates University, Al-Ain, UAE, 15551.}
%\affiliation{Department of physics, United Arab Emirates University, Al-Ain, UAE, 15551.}

%\author{C.-I. Bj{\"o}rnsson} 
%\affiliation{Department of Astronomy, Stockholm University, AlbaNova, SE-106 91 Stockholm, Sweden}

%\author{Francesco Taddia}
%\affiliation{Department of Astronomy, Stockholm University, AlbaNova, SE-106 91 Stockholm, Sweden}

%\author{Peter Lundqvist}
%\affiliation{Department of Astronomy, Stockholm University, AlbaNova, SE-106 91 Stockholm, Sweden}

%\author{Alak K. ray}
%\affiliation{Homi Bhabha Centre for Science Education, 
%Tata Institute of Fundamental Research, Mankhurd, Mumbai, 400088 India}

%\author{Benjamin J. Shappee}
%\affiliation{ Institute for Astronomy, University of Hawai'i, 2680 Woodlawn Drive, Honolulu, HI 96822, USA}

\begin{abstract}
We present the results of Atacama Large Millimeter/submillimeter Array (ALMA) observation in $^{12}$CO(1--0) emission at 0.58 $\times$ 0.52\,pc$^2$ resolution toward the brightest H\,{\sc ii} region N66 of the Small Magellanic Cloud (SMC). The $^{12}$CO(1--0) emission toward the north of N66 reveals the clumpy filaments with multiple velocity components. Our analysis shows that a blueshifted filament at a velocity range 154.4--158.6 km s$^{-1}$ interacts with a redshifted filament at a velocity 158.0--161.8 km s$^{-1}$. A third velocity component in a velocity range 161--165.0\,km\,s$^{-1}$ constitutes hub-filaments. An intermediate-mass young stellar object (YSO) and a young pre-main sequence star cluster have hitherto been reported in the intersection of these filaments. We find a V-shape distribution in the position-velocity diagram at the intersection of two filaments. This indicates the physical association of those filaments due to a cloud-cloud collision. We determine the collision timescale $\sim$\,0.2\,Myr using the relative velocity ($\sim$\,5.1\,km\,s$^{-1}$) and displacement ($\sim$\,1.1\,pc) of those interacting filaments. These results suggest that the event occurred at about 0.2\,Myr ago and triggered the star formation, possibly an intermediate-mass YSO. We report the first observational evidence for a cloud--cloud collision that triggers star formation in N66N of the low metallicity $\sim$0.2\,Z$_{\odot}$ galaxy, the SMC, with similar kinematics as in N159W-South and N159E of the Large Magellanic Cloud.

\end{abstract}
\section{Introduction}
\label{sec:introduction}

The massive stars significantly influence the dynamics and the physical conditions of the interstellar medium and play a key role in galaxy evolution. It is, therefore, essential to understand the fundamental physical processes of the parent molecular cloud. These stars are relatively rare because they are short-lived and evolve quickly compared to their low-mass counterparts. They are often found in clusters, and their early formation phase is highly complex due to the influence on the local environment by gravitational collapse and high radiation pressure \citep{Zinnecker07, Motte18}. The triggered star formation at the shock compressed layers of two colliding clouds is suggested to be a possible way to induce the formation of massive stars in a cluster environment \citep{Inoue13}. Recent advances in molecular line observations have revealed several shreds of evidence for the cloud--cloud collision that triggers the formation of massive cloud cores and stars in the Galactic clouds \citep[e.g.,][]{Dobashi14, Torii15, Torii17, Fukui16, Fukui18a, Fukui18b}. Observations with Atacama Large Millimeter Array (ALMA) have allowed us to extend these findings to the extragalactic clouds. These include N159W-South and N159E in the Large Magellanic Cloud (LMC; \citealt{Fukui15, Tokuda19, Fukui2019}), and some evident star-forming regions in M33 \citep{Tokuda20,Muraoka20,Sano20}.

The Small Magellanic Cloud (SMC) galaxy is an excellent laboratory to study the high-mass star-forming regions in the low-metallicity at a sub-parsec resolution due to its close proximity (61\,kpc; \citealt{Hilditch05}) and low metallicity (0.1--0.2 Z$_{\odot}$; \citealt{Russell92}; \citealt{Rolleston03}; \citealt{Lee05}). The star-forming region N66 is the largest and the most luminous H\,{\sc ii} region in the SMC, which comprises a variety of stellar population, including young stellar objects (YSO), pre-main sequence (PMS) and main-sequence OB star association. The region hosts nearly 33 OB stars, that is about half of the entire SMC hot star population \citep{Massey89, Walborn, Evans06}. The majority of this massive star population is located in the central bar of the nebula that appears as a well-defined arc structure extending from south-east to north-west in an H$\alpha$ map (Figure \ref{fig:IRAC80}). These stars are the major sources of photo-ionization in the central bar of N66 that can effectively trigger star formation via stellar wind and shock \citep{Elmegreen77}. A study of gas and dust content of the N66 by \citet{Rubio2000} shows a tight correlation of H$_2$ emission with CO and infrared aromatic emission. The CO has been largely photo-dissociated by far-ultraviolet (FUV) radiation from nearby massive stars in both central bar and northern filament. Similar characteristic of photo-dissociation regions are reported in many Galactic and extragalactic clouds \citep{Tielens93, roussel07, Naslim15}. N66 is, therefore, the most appropriate target to investigate the high-mass star formation mechanism in the SMC.

The deep imaging survey with the Hubble Space Telescope shows that the region harbors at least five clusters of low-mass PMS stars with a significant age difference \citep{Hennekemper08}. These include the stellar population of age $\leq$5Myr, and those with age $\leq$10Myr. Two PMS star clusters in the north of N66 central bar are younger ($\leq$2.5Myr) than those in the bar, suggesting that these clusters are formed after the central bar population by a different formation mechanism \citep{Hennekemper08}.  \citet{Sewilo13} have reported many high-mass and intermediate-mass YSOs in N66. \citet{Rubio2000} have indicated that the presence of infrared emission peaks both in the central bar and the north of N66 that coincides with the young stellar population identified by \citet{Sewilo13}. A possible sequential star formation due to the photo-dissociation by OB star association in the central bar has been suggested \citep{Rubio2000}. However, the mechanism for forming a relatively younger population in the north of N66 is debated. 

In this letter, we present the first results of ALMA observation of $^{12}$CO(1--0) toward the northern filaments of N66 (N66N) in the SMC. 

%and report a possible cloud--cloud collision that triggers star formation in the northern filaments of N66. 

%We study the kinematics of the velocity structures, physical properties of molecular cores, and their possible association with the young stellar population for a better understanding of star formation mechanism away from the central bar. In section 3, we present our results on kinematic studies, filament and molecular core properties. In section 5 we propose the cloud-cloud collision as the most likely mechanism for star formation in N66-N.

%YSOs and The infrared spectroscopic observations have shown compact H2 emission knots, and emission peaks coincident with the ionized gas revealing a nice PDR (Rubio et al. 2000). The massive YSOs in N66 have been extensively studied by Rubio et al (2018), hence our results on clump properties can be directly compared. 

%\section{HII region N\,44}

\section{Observations/Data}
We used the ALMA archival data (P.I., Erik Muller, \#2015.1.01296.S) of the N66N. The Band~3 (3\,mm) observations were performed using the 12\,m array with the configurations of C36-1/2, C36-2/3, and C40-6 during the Cycle 3 and 4 observing seasons. The representative spectral window targeted the $^{12}$CO($J$ = 1--0) line with a frequency resolution of 61\,kHz and a channel number of 3840. We performed the imaging process using the Common Astronomy Software Application (CASA) package \citep{McMullin07} version 5.4.1. The weighting scheme of the \texttt{tclean} was $``$Briggs$"$ with a robust parameter of 0.5. The \texttt{auto-multithresh} procedure \citep{Kepley20} automatically selected the emission mask in the dirty and residual images in \texttt{tclean}. We continued the deconvolution process until the intensity of the residual image attains the $\sim$1$\sigma$ noise level. The resultant beam size and rms noise level are 2\farcs0\,$\times$\,1\farcs8 (P.A. = 10.0\,deg; 0.58\,$\times$\,0.52\,pc$^2$) and 0.022\,Jy\,beam$^{-1}$ (=0.59\,K) at a velocity resolution of 0.2\,km\,s$^{-1}$, respectively.

To investigate the missing flux of the 12\,m array data, we retrieved another ALMA program (P.I., Claudia, Agliozzo, \#2017.A.00054.S), which contains the Atacama Compact Array (ACA) 7\,m array data alone in $^{12}$CO($J$ = 1--0) toward the N66 region. We applied the same flow described in the previous paragraph to the data reduction (imaging) process. We converted the 12\,m array data to be the same resolution as the 7\,m array data, and then produced the integrated intensity ratio map in the pixels at more than 3$\sigma$ detection. Since the flux ratio between the two data set is almost 1, we conclude that the 12\,m array data does not suffer from the significant missing flux. We used the 12\,m array alone data throughout this manuscript.

\section{Results}
\subsection{$^{12}${\rm CO(1--0)} spatial distributions and filaments}
\begin{figure*}
%    \centering
%    \includegraphics[scale=0.4]{IRAC80_yso2.eps}
%    \includegraphics[trim=0.2 0 0.9 0, clip, width=0.5\textwidth]{N66_OB_YSO_SNR3.eps}
\includegraphics[trim=0.2 0 0.9 0, clip, width=0.5\textwidth]{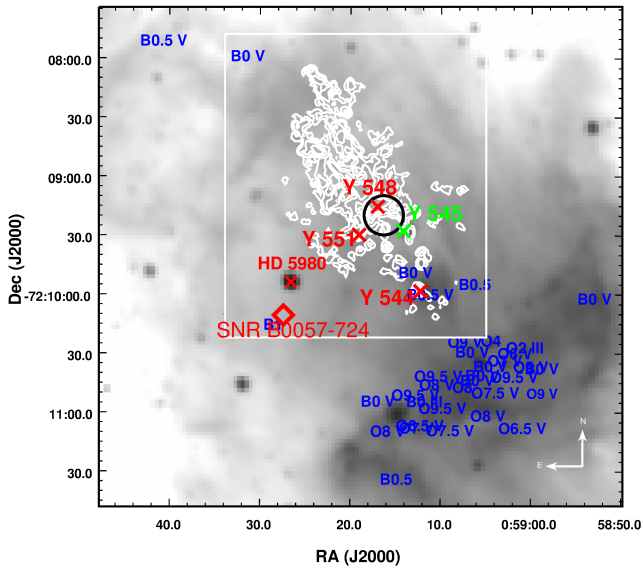}
\includegraphics[trim=1.2 0.2 0.4 0, clip, width=0.4\textwidth]{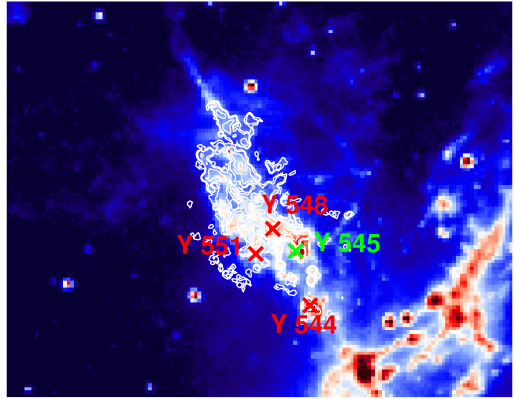}
    \hspace{-0.6 cm}
%    \vspace{-0.6 cm}
    \caption{Left: H$\alpha$ map of N66 is shown along with $^{12}$CO(1--0) emission (contour) of N66N. The white box shows the region N66N of this study. The OB stars, SNR B0057-724, and 4 YSOs close to the $^{12}$CO(1--0) filaments are labeled. The location of PMS star cluster 2 \citep{Hennekemper08} is shown (black circle) for comparison. Right: Spitzer 8.0$\micron$ map of N66N is shown along with $^{12}$CO(1--0) emission in contour.}
    \label{fig:IRAC80}
\end{figure*}

\begin{figure}
    \centering
    \includegraphics[width=0.45\textwidth]{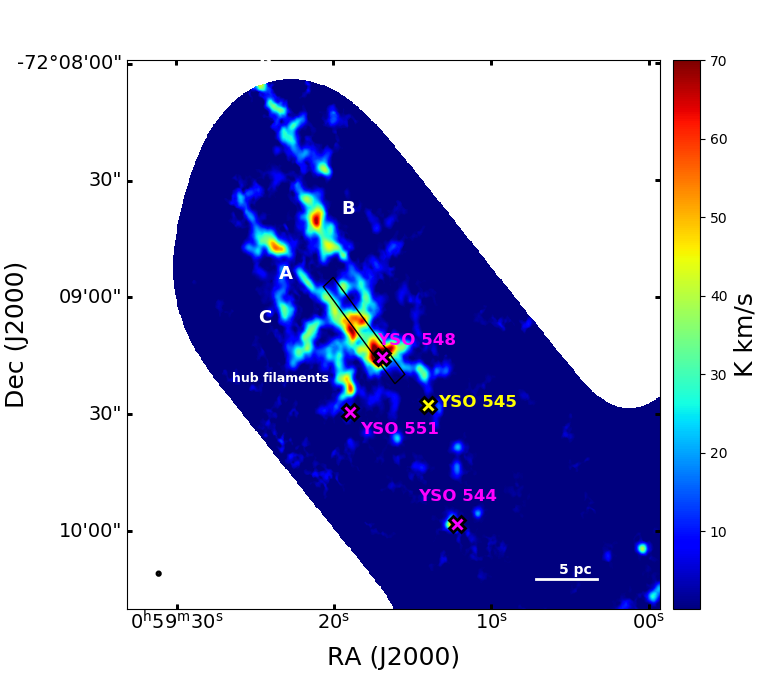}
    \caption{The $^{12}$CO(1--0) integrated intensity map of N66N. The filaments A, B, C, and the positions of YSOs are labeled for comparison. The angular resolution 2\farcs0 $\times$ 1\farcs8 is shown (black ellipse) in the left bottom corner}
    \label{moment0}
\end{figure}

Figure \ref{moment0} shows the velocity integrated intensity distribution of $^{12}$CO(1--0) emission, which reveals the filamentary and clumpy structures of N66N. Two well-ordered filaments (A and B) appear as elongated structures extending from the dense clumps toward the north-east of N66 (Figure \ref{moment0}). There is a third filamentary structure which constitutes multiple small filaments, similar to the structures indicated as hub-filaments in many other high-mass star-forming regions \citep{Peretto13, Motte18, Tokuda19, Kumar20}. In Figure \ref{fig:IRAC80}, we show the distribution of $^{12}$CO(1--0) emission in N66N on an H$\alpha$ map \citep{smith98}, and the Spitzer 8.0$\micron$ map \citep{Gordon11}. For comparison, we show the location of 4 YSOs closer to the N66N and 41 OB stars \citep{Dufton19} that distribute over the entire N66. The ionized gas traced by H$\alpha$ emission shows a giant H\,{\sc ii} region in the center of N66 that appears as a bar. Toward the N66N filamentary complex, the H$\alpha$ emission appears to be more diffuse. At the south-west of the $^{12}$CO(1--0) filaments, there exists a compact H$\alpha$ emission complex. The distribution of 8.0$\micron$ emission in N66N shows multiple filaments, and the morphology resembles the spatial distribution of $^{12}$CO(1--0) emission. 

%This emission morphology suggests the presence of colliding molecular clouds at the region where two filaments join. 

%To obtain a further indication of the cloud-cloud collision, we inspect the velocity structures of this region in section 3.2.

To estimate the filament's mass and size, we identify the filament structures using the python package \textbf{astrodendro} \citep{Rosolowsky08}. \textbf{Astrodendro} characterizes the molecular gas as a structure tree with leaves, branches, and trunks in a three-dimensional data cube. The trunk represents the parent cloud, which constitutes the brightest structures as leaves, and the low-density connecting structures as branches. We identify the filaments as the parent structures that are the trunks in the dendrogram. \textbf{Astrodendro} provides basic parameters such as size, velocity dispersion, and flux of molecular structures. We obtain the length and width of the filaments, A and B, from the respective major and minor axes of the trunks ($\sigma_x$, $\sigma_y$). We derive the apparent velocity width of each filament using the velocity dispersion, $\sigma_v$, assuming a Gaussian distribution ($\Delta V$=$2\sqrt{2ln2}\sigma_v$). The derived velocity dispersion of filament A is 0.7\,km\,s$^{-1}$ that corresponds to a velocity width of 1.7 km s$^{-1}$. The length and width of filament A is 13\,pc, and 3.8\,pc respectively. Assuming a CO-to-H$_2$ conversion factor 7.5$\times$10$^{20}$cm$^{-2}$ (K\,km\,s$^{-1}$)$^{-1}$ for the SMC \citep{Muraoka17}, we find an H$_2$ column density of 3.2$\times 10^{22}$\,cm$^{-2}$ for the filament A that translates to an average mass $\sim$\,2.3$\times$10$^{3}$\,M$_{\odot}$. The filament B shows a velocity dispersion of 0.88 km s$^{-1}$ ($\Delta V$=2.1\,km\,s$^{-1}$) with a length $\sim$\,21\,pc and width $\sim$\,3\,pc. The H$_2$ column density of filament B is estimated to be 5.8$\times$10$^{22}$cm$^{-2}$ with a mass $\sim$\,4.3$\times$10$^{3}$\,M$_{\odot}$.

%In order to determine the size and mass of the filaments we consider thof these filaments using 12CO(1-0) emission assuming the Xco factor in the SMC, 2.0$\times$10$^{20}$cm$^{-2}$ (K km s$^{-1}$)$^{-1}$. The total mass of the cloud is estimated to be $\sim$.. $M_{\odot}$. The filament 1 has a mass of $\sim$..$M_{\odot}$, FWHM line width of $\sim$.. and length .. pc, and filament 2 shows a FWHM line width of $\sim$..., mass $\sim$..$M_{\odot}$ and length .. pc. 

%Figure \ref{fig:IRAC80}a shows the comparison of 12CO emission with the IRAC 8.0$\micron$m map obtained as part of SAGE observation. IRAC 8.0 micron emission traces Poly cyclc aromatic hydrocarbon emission. The strucute of filaments in IRAC 8.0 micron emission resembles with 12CO emission. Figure \ref{fig:IRAC80}a shows a comparison of CO emission with H$\alpha$ emission. There is a bright HII region at the south-eastern edge of filaments. 

%\section{12CO filaments}
%\subsection{Extinction correction}
%{\blue dust extinction and reddening correction from Halpha to Hbeta ratio}
\begin{figure*}
    \centering
\includegraphics[width=0.88\textwidth]{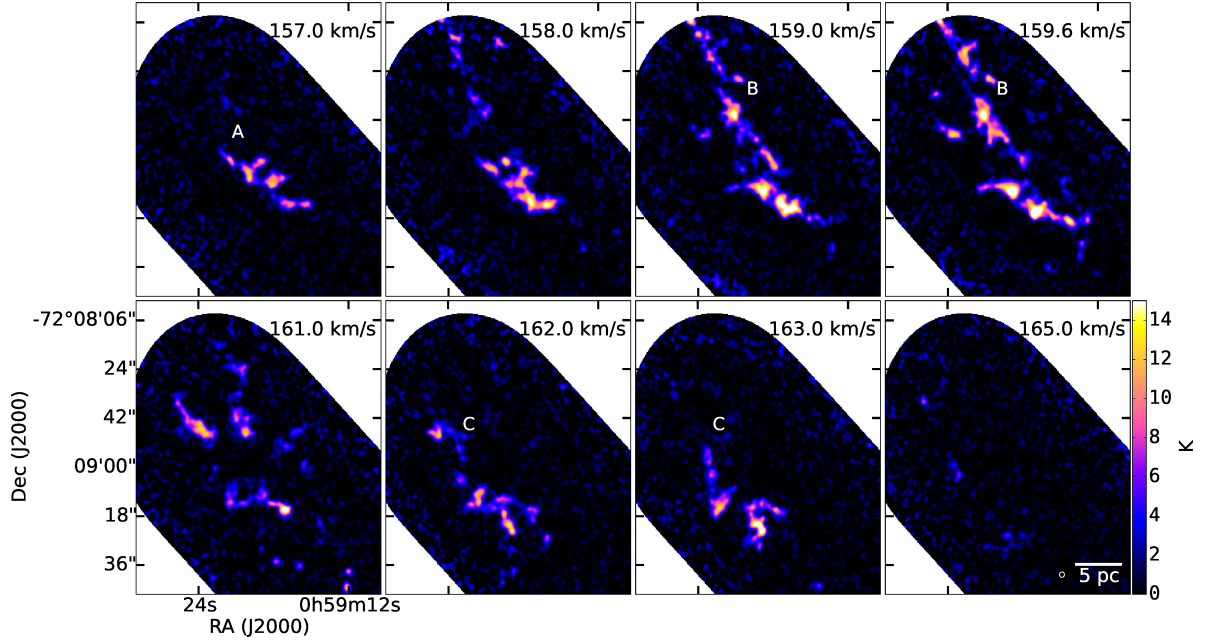}
    \caption{The $^{12}$CO(1--0) velocity channel maps toward the N66N. The velocity is given in the upper right corner of each panel. The angular resolution 2\farcs0 $\times$ 1\farcs8 is shown in the right bottom corner. }
    \label{channelmap}
\end{figure*}

\begin{figure*}
% \centering
\includegraphics[width=0.5\textwidth]{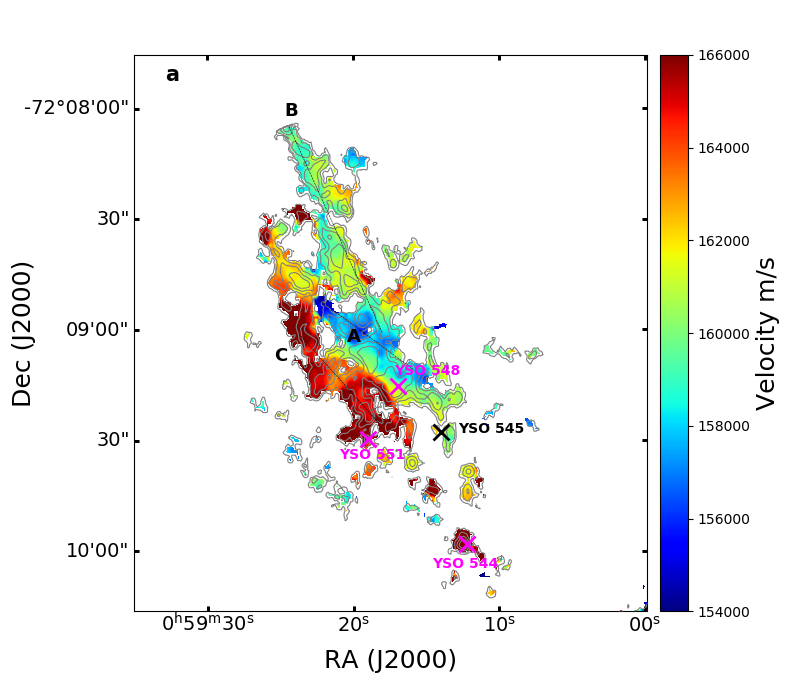}
\hspace{0.5cm}
\includegraphics[width=0.418\textwidth]{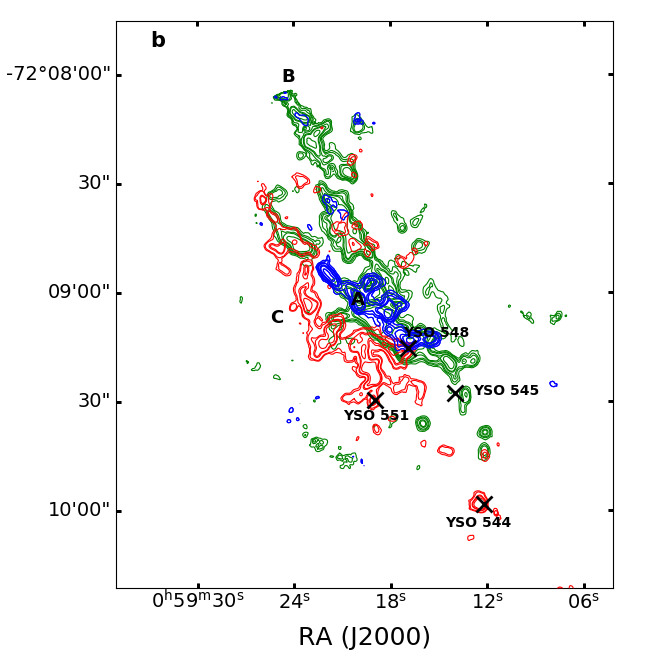}
%\vspace{1.0cm}
    \caption{a: The $^{12}$CO(1--0) first-moment intensity weighted velocity map of N66N. b: The blueshifted component A (blue), the second velocity component B (green), and the third redshifted component C (red) are shown in $^{12}$CO(1--0) integrated intensity contours. The contour levels are 3, 20, 30, and 60\,K\,km\,s$^{-1}$ }
    \label{moment1}
\end{figure*}

%\begin{figure}
% \centering
%    \includegraphics[scale=0.7]{Figure_1.png}\\
%    \includegraphics[scale=0.3]{N66_PV_filament1_blue_latest.eps}
%    \includegraphics[scale=0.3]{N66_PV_filament2_blue_latest.eps}
%    \vspace{-0.5cm}
%    \includegraphics[scale=0.3]{N66_PV_filament3_redlong_latest.eps}
%    \includegraphics[scale=0.45]{N66_PV_filament_join_V.eps}
%    \caption{The position-velocity diagrams of velocity components A, B and C (top to bottom) along the cuts (black dotted lines) shown in Fig\ref{moment1}}
%    \label{PV1}
%\end{figure}

\begin{figure*}
\centering
\includegraphics[scale=0.4]{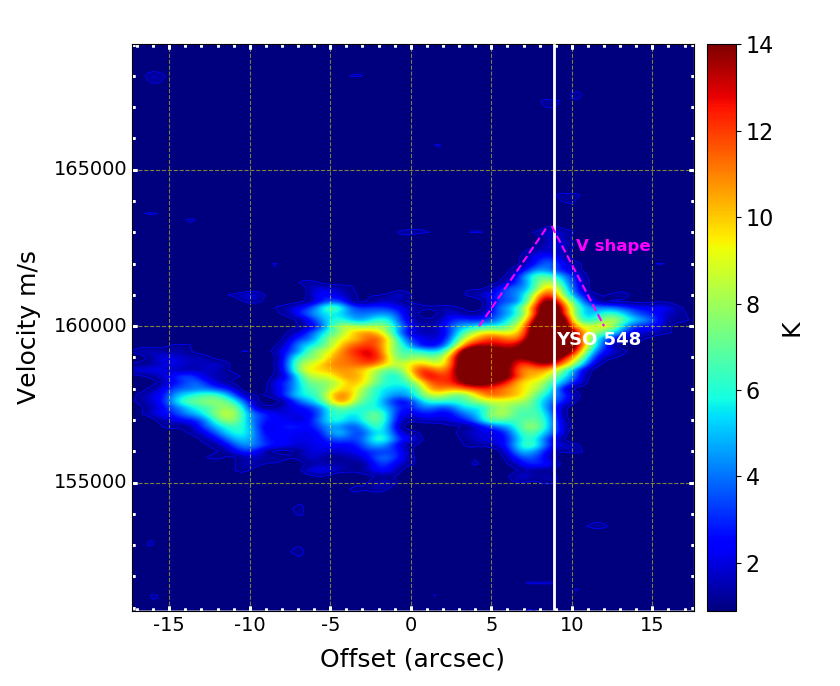}
    \caption{The V-shape distribution in a position-velocity diagram at the intersection of filaments A and B across the cut shown in Figure \ref{moment0}. The position of YSO\,548 along the V-shape is labeled.}
    \label{PV2}
\end{figure*}

\subsection{Velocity structures}

The complexity of velocity structures is visible in a series of channel maps (Figure \ref{channelmap}). The molecular cloud has an elongated filamentary structure that spread over a velocity range 154.4--165.2\,km\,s$^{-1}$. Filament A shows a velocity range 154.4--158.6 km s$^{-1}$ and filament B shows a velocity range 158.0--161.8\,km s$^{-1}$. We find a third velocity component (Filament C) in a velocity range 161--165.0\,km\,s$^{-1}$. 

%This third velocity component constitutes the hub-filaments (Figure \ref{moment0}). 

%We inspect the first-moment intensity weighted map and the position-velocity (PV) diagram to investigate further key evidence for a cloud--cloud collision.

Figure \ref{moment1}a shows the first-moment intensity weighted velocity (moment\,1) map of $^{12}$CO(1--0) emission in N66N. The moment\,1 map reveals the blueshifted and redshifted velocity components (A and B) relative to the systemic velocity $\sim$158\,km\,s$^{-1}$.
 In Figure \ref{moment1}b, we show the three velocity components in $^{12}$CO(1--0) integrated intensity contours. The blue contour represents the blueshifted filament A with the peak emission at a velocity of 157.4\,km\,s$^{-1}$ and the green contour represents the redshifted filament B with a velocity peak at 160\,km\,s$^{-1}$. The red represents the redshifted third component C at a peak velocity of 163\,km\,s$^{-1}$.

%In Figure \ref{PV1} we show
%the position velocity (PV) diagrams of these Filaments A, B and C
%along the cuts (black lines) shown in Figure \ref{moment1}. The PV
%diagrams show one blue-shifted and two-red shifted velocity components
%with respect to the systemic velocity $\sim$158 km s$^{-1}$. The
%blue-shifted filament (A) has a peak emission at a velocity of 157.4
%km s$^{-1}$. The red-shifted filament B shows a peak 12CO(1-0)
%emission at a velocity 160 km s$^{-1}$. The hub-filaments and the
%third component (C) show a peak emission at a velocity 163km
%s$^{-1}$ that appear as a second red-shifted velocity component in
%south-east of the blue-shifted filament A in Figure \ref{moment1}. 

The position-velocity (PV) diagram at the intersection of filaments A and B shows a V-shape gas distribution (Figure \ref{PV2}), indicating the physical connection of two filaments. This V-shape gas distribution at the filament intersection in the PV diagram is a key observational signature of the cloud--cloud collision (discussion in section 5).

%In addition a bridging feature is noticed at the
%intersection of two filaments at a velocity of $\sim$158.8 km
%s$^{-1}$. The two filaments are physically connected by a bridge of
%diffuse emission in PV diagram (Figure \ref{PV1}). The
%complementary distribution of filaments, bridge feature and V shape
%gas distribution at the filament intersection in PV diagrams are the
%key observational signatures of a cloud-cloud collision. 

%The filaments
%with complementary velocity components, V-shape and bridge features inPV diagrams have been reported in many high-mass star forming regionsof the Milky Way (Ref). The authors interpret these features as aresult of the cloud-cloud collision that trigger the formation ofhigh-mass stars.  

\section{CO cores in filaments and associated Young stellar population}
The velocity integrated $^{12}$CO(1--0) intensity map of N66N shows localized emission peaks along the filaments. We use \textbf{astrodendro} to identify the molecular cores and determine their virial masses, following the same method as in \citet{Naslim18, nayana20}. We identify the molecular cores as the dendrogram leaves, which are the brightest and smallest structures representing the top level of the dendrogram. We detect three CO cores along the filament A which show the virial masses 19, 350, and 1300\,M$_{\odot}$, and the total of nine CO cores along the filament B in a mass range 120-565\,M$_{\odot}$. There are 16 CO cores in the third velocity component C in a mass range 20-225\,M$_{\odot}$. We note three molecular cores in the intersection of A and B with the virial masses 156, 173 and 1442\,M$_{\odot}$. 

%In Figure \ref{moment1}b, we show the $^{12}$CO(1--0) second-moment map of N66N that indicates the velocity dispersion along the filaments. The CO cores in the filament intersection show relatively larger velocity dispersion. 

%The mass of each CO cores are determined by assuming the cores with spherical shape and a truncated density profile, $\rho \sim$r$^{-1}$, and we derive the virial mass $M_{vir}$ (in $M_{\odot}$) using the formula, 
%A dense clump at the intersection is a direct parental core of YSO 548
%\begin{equation}
% M_{vir}=209R(\Delta V)^2
%\end{equation}

%where R is the radius of CO core in pc, $\Delta V$ is the velocity width in km s$^{-1}$. 
%an X$_{co}$ factor of 2.0$\times$10$^{20}$cm$^{-2}$ (K km s$^{-1}$)$^{-1}$.
%\begin{figure}
% \centering
%    \includegraphics[trim=28 0.0 90 0.0, clip, scale=0.8]{N66_molecular_cores.eps}
%    \includegraphics[trim=90 0.0 50 0.0,clip, scale=0.6]{N66_moment2_3filaments.png}
%    \label{leaves}
%end{figure}

%\begin{figure*}
%\centering
%    \includegraphics[trim=28 0.0 80 0.0, clip, scale=0.56]{N66_molecular_cores.eps}
%    \includegraphics[trim=90 0.0 50 0.0,clip, scale=0.42]{N66_moment2_3filaments.png}
%    \caption{Left: Molecular cores identified by astrodendro are plotted on a velocity integrated intensity map of 12CO(1-0). Right: The second moment intensity weighted velocity map of 12CO(1-0) that indicates the velocity dispersion.}
%    \label{leaves}
%\end{figure*}
%\subsection{Associated Young Stellar Objects}
To investigate the ongoing star formation in N66N, we adopt the YSOs identified by \citet{Sewilo13}. Figures \ref{moment0} and \ref{moment1} show the positions of three high-reliable YSOs (548, 551, 544) and a less-reliable YSO (545). 

%The YSO\,548 is located in the intersection of filaments A and B. YSO\,551 is at the hub-filament in the south-east. YSO\,544 is located away from the filaments in an H$\alpha$ complex at the south-west of the filament intersection. 

%The less reliable YSO\,545 is located slightly away from the filament intersection. 

%Figure\ref{leaves}, we compare the position of CO cores and their velocity dispersion with the associated youmg stellar population. YSOs 548, 551 and 544 are associated with the CO cores 27, 28 and 29 (Table\ref{tab:clump-details-astrodendro}).

\section{Discussion}

%The ALMA observation of $^{12}$CO(1--0) emission in N66N reveals filamentary and clumpy nature of molecular cloud. 
Observations of molecular clouds in the Milky Way and the LMC have shown that filaments of different velocities can be a result of colliding clouds triggering high-mass star formation. To find proper evidence for a cloud--cloud collision, we use three methods discussed by \citet{Fukui18a}. a) Velocity distribution of molecular gas in channel maps: we find multiple velocity components in channel maps. One filament at a blueshifted velocity gradient 154.4--158.6\,km\,s$^{-1}$ (filament A) seems to interact with a redshifted filament of velocity 158.0--161.8\,km\,s$^{-1}$ (filament B). Multiple hub-filaments are found in a third redshifted component at a velocity range 161--165.0\,km\,s$^{-1}$. b) Velocity distribution in a first-moment map: three velocity components are found. A YSO has been reported at the intersection \citep{Sewilo13}. A similar filament distribution is reported in N159E of the LMC \citep{Saigo17, Fukui2019}. c) Position-velocity diagram: we find a V-shape gas distribution at the intersection of filaments A and B, indicating their physical interaction. These observational signatures are consistent with the cloud--cloud collision reported in N159W-South, N159E of the LMC, and NGC 604 of M33 \citep{Tokuda19, Fukui2019, Muraoka20}.
%These observational signatures are consistent with the cloud--cloud collision reported in many star-forming regions in the Milky Way such as M42, M43 of the Orion nebula cluster, RCW38, Westerlund 2, and NGC 3603, N159W-South and N159E of the LMC \citep{Fukui16, Fukui18a, fukui14, Furukawa09, Tokuda19}.

%The high-spatial-resolution $^{12}$CO(1--0) map of N66N in the SMC gives us a unique opportunity to study the star formation at a lower metallicity (0.1--0.2\,Z$_{\odot}$). 
 N159W-South region of the LMC shows complex filaments with multiple velocities and a V-shape distribution in the PV diagram \citep{Tokuda19}. The formation of such complicated filamentary structures and the massive YSOs at the intersection is suggested to be due to the large-scale colliding flow. At a distance of 50$\sim$pc away from N159W-South, the N159E star-forming region shows similar filamentary characteristics with multiple velocity components and a massive YSO at the intersection \citep{Saigo17}. Using a higher spatial-resolution observation, \citet{Fukui2019} reveal more than a few tens of filamentary structures and the protostar activities at the filament intersection. The orientation of these multiple filaments follows a large-scale H\,{\sc i} flow, which is driven by the tidal interaction between the SMC and the LMC. Comparing these observations with the magnetohydrodynamic (MHD) numerical simulation \citep{inoue18}, \citet{Fukui2019} suggest that the formation of filaments and protostars be a result of H\,{\sc i} colliding flow. \citet{Muraoka20} claim a similar large scale colliding flow as a potential cause for the formation of multiple filaments and protostars in the NGC 604 cloud of M33. \citet{McClure-Griffiths18} report the large scale H\,{\sc i} outflows extending about 2 kiloparsecs from the SMC bar, which can be driven by the tidal/ram pressure interaction between the SMC and the LMC. The collision in N66N may be a part of this H\,{\sc i} outflow. \citet{McClure-Griffiths18} do not provide any evidence for the return of this H\,{\sc i} flow into the SMC. To further explain whether the collision in N66N is related to a large-scale flow, we need to study the velocity structure of H\,{\sc i} gas.

%\citet{Fukui18a} have reported $^{12}$CO(1--0) emission with multiple velocity components and the PV diagram with V-shape gas distribution toward the H\,{\sc ii} regions M42 and M43 of the Milky Way. The authors interpret this result that two clouds collided with each other at a supersonic velocity to trigger the formation of massive stars in M42 and M43. The synthetic observations of the cloud--cloud collision have shown that a V-shape distribution is found if the relative motion of two colliding clouds has an angle of 45$^{\circ}$ or 0$^{\circ}$ to the line of sight \citep{Fukui18a}. 

%The V shape gas distribution in the PV diagram is found in many other interacting clouds, such as RCW 38, M20, and NGC 3603

The MHD numerical simulation of two colliding clouds of different sizes have shown the formation of a shock-compressed layer between two clouds \citep{Inoue13}. The density and turbulent velocity are enhanced in the shock-compressed interface due to an amplified magnetic field, and the magneto-hydrodynamical fluid flow induces the formation of filaments in multiple velocities \citep{Inoue13}. The MHD simulations by \citet{inoue18} claim that the massive filaments are formed in the shock compressed layer between the cloud and a large scale gas flow. If the actual line mass exceeds the critical value, the filaments become gravitationally unstable, which further collapse into fragments and form molecular cores. The mass accretion creates multiple dense cores at the intersection, where the most massive core becomes an O star accompanied by several low-mass stars. Since the colliding cloud is very massive ($>$50\,M$_{\odot}$), the mass accretion rate on to the central cloud can be as large as $\sim$10$^{-4}$\,M$_{\odot}$yr$^{-1}$. This leads to the formation of massive cores as large as $\sim$100\,M$_{\odot}$ in 0.3\,Myr \citep{Inoue13, Fukui16}. We estimate the timescale of cloud collision in N66N using the relative displacement and velocity of two interacting filaments. The displacement between two filaments on their relative location at the point of intersection is estimated to be 0.8\,pc with a relative velocity of 3.6 km s$^{-1}$. If we assume the relative motion of colliding clouds has an angle of 45$^{\circ}$ to the line of sight, we get the projection corrected velocity $\sim$\,5.1\,km\,s$^{-1}$ and the displacement $\sim$\,1.1\,pc. We estimate the collision time scale to be $\sim$0.2\,Myr. Three $^{12}$CO(1--0) cores in a mass range 156--1442\,M$_{\odot}$ are identified at the filament intersection. High column density is an essential initial condition for forming high-mass stars by cloud--cloud collision. A peak column density of $\sim$10$^{23}$cm$^{-2}$ can produce nearly 10 O stars, while a single O star can be produced with a peak column density of $\sim$10$^{22}$ cm$^{-2}$ \citep{Fukui16}.  The H$_2$ column densities of two interacting filaments in N66N are 3.2$\times 10^{22}$\,cm$^{-2}$ and 5.8$\times 10^{22}$\,cm$^{-2}$.

The young stellar population and massive star content of N66 have been extensively studied \citep{Sewilo13, Massey89}. To investigate the ongoing star formation, we compare the positions of YSOs \citep{Sewilo13}, and OB stars \citep{Dufton19} with the CO cores in filaments. We find that majority of OB stars and YSOs are in the N66 central bar, and four YSOs (548, 551, 544, 545) are located in the north-east. Among these, the most reliable YSO\,548 is at the filament intersection and YSO\,551 is in a hub-filament. The YSO\,544 is in a compact H\,{\sc ii} region between the filament intersection and the central bar. We do not have enough parametric information such as the mass and luminosity of YSO\,548. \citet{Sewilo13} provide an 8.0$\micron$ magnitude of [10.69] for YSO\,548, indicating an intermediate-mass (mass range 5-10\,M$_{\odot}$) object according to the selection criterion suggested by \citet{Chen09}. The YSO\,548 is associated with a $^{12}$CO core of virial mass 156\,M$_{\odot}$. It is also possible that a massive YSO remains to be formed in the densest core at the intersection. \citet{Sewilo13} report a mass of $\sim$10.1M$_{\odot}$ and a luminosity of 3.24$\times$10$^{3}$L$_{\odot}$ for YSO\,544. The $^{12}$CO core associated with YSO\,544 shows a virial mass of $\sim$398M$_{\odot}$, although there is no extended CO emission. 

The $^{12}$CO(1--0) filaments in N66N have similar morphology with the 8.0$\micron$ emission, indicating the photoelectric heating of polycyclic aromatic hydrocarbon by FUV radiation, possibly from OB stars in the central bar. The extended low-density H$\alpha$ emission and $^{12}$CO(1--0) filaments in N66N show a vertical orientation with the H\,{\sc ii} region in the central bar. \citet{Ye91} suggest a high-velocity 'Champagne flow' from the exciting OB stars in the central bar toward the north-east. The H\,{\sc ii} region in the central bar are photo-dissociating the natal molecular clouds, but some of the clouds are still surviving at N66N where the cloud collision seems to be taking place, resulting in a formation of at least an intermediate-mass star at the intersection of two filaments. In addition to the YSO\,548, a PMS star cluster is located in the filament intersection. This PMS star cluster is a relatively younger population (0.25--2.5\,Myr) than those in the central bar (0.5--10\,Myr; \citealt{Hennekemper08}) and located 24\,pc away from the central bar in the filament intersection. We expect this low-mass PMS cluster has been formed prior to the cloud--cloud collision as it is found to be slightly older than the estimated collision timescale $\sim$0.2\,Myr. \citet{Gouliermis08} have proposed, this PMS population outside of the central bar can be triggered by the wind-driven expanding H\,{\sc ii} region blown by the supernova remnant SNR B0057-724 \citep{Ye91,Naze02} (Figure 1), which is in the east of the filament intersection at a projected distance of $\sim$21\,pc. Although the redshifted filament C has an overlap with an edge of the X-ray emission \citep{Naze02}, we do not find a continuous velocity gradient with an expanding shell of molecular gas that is blown from SNR\,B0057-724 in the direction of filament intersection. 

%at a projected distance of 0.4\,pc from the point where two filaments join

We finally remark on star formation in the SMC-like low-metallicity environment (0.1--0.2\,Z$_{\odot}$) and the future prospect of this study. The numerical simulations by \citet{Ricotti97, Ricotti02} show that the kinetic energy dissipation of the cloud collision decreases in a lower-metallicity condition due to the longer cooling time-scale of the shocked gas than the characteristic collision time; hence a significant effect on the star formation is expected. Our finding indicates that cloud--cloud collision does occur and work as a trigger of star formation in low-metallicity environment. However, the N66N region hosts several YSOs, such as YSO\,544, associated with a compact/isolated CO clump, which is not evident in more than two velocity components at the current angular resolution. The subsequent step is to understand how common the cloud--cloud collision throughout the N66 region and the SMC, and how the properties of parental molecular clouds affect the mass and luminosity of embedded YSO. A more comprehensive analysis extending to the other YSOs using the current data set in N66 will be presented in a future paper.

%YSO 544 stays away from the filament intersection toward the south, within a compact HII region along with the PMS cluster 3 \citep{Hennekemper08} at a distance of 19pc from SNR B0057-724 and 15pc from the central OB star association. Our studies of 12CO(1-0) velocity structures clearly show evidences for a cloud-cloud collision at the northern filament intersection, hence we suggest this scenario can be the main triggering source for the formation of young star population in this region.

\section{Conclusion}

We report clumpy filaments with multiple velocity components toward N66N in the SMC by ALMA. Our results are concluded as follow:

%Based on the observation, we suggest a cloud-cloud collision that likely triggers star formation in N66N. Our results are concluded as follow:

\begin{enumerate}
\item The ALMA observation of $^{12}$CO(1--0) emission in N66N shows a blueshifted velocity component (A) in a velocity range 154.4--158.6\,km\,s$^{-1}$ and a redshifted component (B) in a velocity 158.0--161.8\,km\,s$^{-1}$. A third redshifted component (C) in a velocity 161-165.0\,km\,s$^{-1}$ shows hub-filament distribution. 
\item An intermediate-mass YSO has been found at the intersection of filaments A and B. We find a V-shape gas distribution in the PV diagram taken at the intersection of filaments A and B, indicating their physical association. These filament characteristics are similar to the cloud--cloud collision reported in N159E of the LMC. 

\item We find the H$_2$ column densities of two interacting filaments using a CO-to-H$_2$ conversion factor 7.5$\times$10$^{20}$cm$^{-2}$ (K\,km\,s$^{-1}$)$^{-1}$. The redshifted component has an H$_2$ column density of 5.8$\times 10^{22}$\,cm$^{-2}$ and mass $\sim$4.3$\times$10$^{3}$\,M$_{\odot}$, while the blueshifted component shows a column density 3.2$\times 10^{22}$\,cm$^{-2}$ and mass $\sim$\,2.3$\times$10$^{3}$\,M$_{\odot}$.   
\item We estimate the timescale of collision $\sim$\,0.2\,Myr using the relative velocity $\sim$\,5.1\,km\,s$^{-1}$ and displacement $\sim$\,1.1\,pc of two interacting filaments.

\item Comparing our findings with the results of MHD simulations \citep{inoue18}, molecular line observations in N159W-South and N159E of the LMC \citep{Fukui2019} and the NGC 604 of M33 \citep{Muraoka20}, a cloud collision is a possible triggering mechanism for star formation in N66N. Our results suggest that the collision occurred $\sim$\,0.2\,Myr ago, resulting the formation of filaments with multiple velocity components and triggered the formation of an intermediate-mass YSO.

%\item The PV diagrams show V shape gas districution and bridge feaures to support the possibilities for a cloud-cloud collision.
%\item 

%where all the velocity components join in a dense clump. The morphology of the filamentary structures and the v-shaped velocity components in PV diagram resemble well with the cloud-cloud collision model of high-star formation. 

\end{enumerate}

\section*{Acknowledgments}

This paper makes use of the following ALMA data:
ADS/JAO.ALMA\,$\#$2015.1.01296.S, 2017.A.00054.S. ALMA is a partnership of the ESO, NSF,
NINS, NRC, NSC, and ASIAA. The Joint ALMA Observatory
is operated by the ESO, AUI/NRAO, and NAOJ.
This research has been supported by United Arab Emirates University, under start-up grant 31S378 and UPAR grant G00003479. 

\bibliography{N66CO_filament}{}
\bibliographystyle{aasjournal}

%\clearpage

\end{document}